# Depth-resolved vascular profile features for artery-vein classification in OCT and OCT angiography of human retina

**TOBILOBA ADEJUMO,**[1,3] **TAE-HOON KIM,**[1,3] **DAVID LE,**[1] **TAEYOON SON,**[1] **GUANGYING MA,**[1] **AND XINCHENG YAO**[1,2,*]

[1] *Department of Biomedical Engineering, University of Illinois at Chicago, Chicago, IL 60607, USA*
[2] *Department of Ophthalmology and Visual Sciences, University of Illinois at Chicago, Chicago, IL 60612, USA*
[3] *These authors contributed equally to this work*
[*] *xcy@uic.edu*

**Abstract:** This study is to characterize reflectance profiles of retinal blood vessels in optical coherence tomography (OCT), and to test the potential of using these vascular features to guide artery-vein classification in OCT angiography (OCTA) of the human retina. Depth-resolved OCT reveals unique features of retinal arteries and veins. Retinal arteries show hyper-reflective boundaries at both upper (inner side towards the vitreous) and lower (outer side towards the choroid) walls. In contrast, retinal veins reveal hyper-reflectivity at the upper boundary only. Uniform lumen intensity was observed in both small and large arteries. However, the venous lumen intensity was dependent on the vessel size. Small veins exhibit a hyper-reflective zone at the bottom half of the lumen, while large veins show a hypo-reflective zone at the bottom half of the lumen.

## 1. Introduction

Systemic diseases are known to frequently target the retinal neurovascular system. For example, reduced retinal blood flow velocities in the short posterior ciliary artery might reflect glaucoma development [1]. Retinal arterial narrowing [2, 3] and venous beading [4, 5] are common features in diabetic retinopathy (DR). Retinal venous dilation has been also reported in diabetic patients [6, 7]. Moreover, retinal venous narrowing and flow reduction were observed in Alzheimer's patients [8], and retinal venous abnormality was detected in Parkinson's disease [9]. Because different diseases are affecting the retinal arteries and veins in different ways, differential artery-vein (AV) analysis promises improved sensitivity for disease prognosis and staging classification in various eye conditions and systemic diseases. Thus, differential AV analysis has recently been explored for retinal disease detection and staging classification, and AV caliber and tortuosity ratio have been verified as useful predictors of retinal diseases [10].

To achieve differential AV analysis, AV classification is a prerequisite. To date, several methods have been reported for the AV classification in color fundus photography [11-13] and fluorescein angiography [14]. However, these imaging modalities do not provide information on layer-specific vascular pathology. As a new modality of optical coherence tomography (OCT), OCT angiography (OCTA) enables noninvasive label-free imaging of retinal vasculature at capillary level resolution [15-18]. OCTA has been quickly adopted in lab research and clinical management of various diseases [19] such as DR [20], sickle cell retinopathy[21], Alzheimer's disease [22], glaucoma [23], retinal vein occlusion [24], age-related macular degeneration (AMD) [25]. However, clinical OCTA does not provide the function of AV classification.

Several strategies have been explored for AV classification in OCTA [26]. Xu et al. demonstrated AV classification by identifying deep capillary plexus (DCP) vortices from OCTA scans [27]. Alam et al. established AV classification in OCTA image by a color fundus image guided approach [28]. Alam et al. also showed the use of OCT feature analysis to aid AV classification in OCTA [29]. Son et al. demonstrated the feasibility of using near-infrared OCT oximetry to aid AV classification in OCTA [30]. Local features on *en face* retinal images can be also used for AV classification. As an example, veins are relatively darker than arteries; arteries are narrower than the veins around them. Also, the central reflex of retinal arteries is brighter than that of veins [31]. These characteristic features can discriminate arteries and veins based on size, color, and topological properties. Recently, a fully convolutional network AV-Net has been also demonstrated for automated AV classification in OCTA [32]. The AV-Net is based on deep learning processing of *en face* OCT and OCTA images. The *en face* OCT/OCTA images can only provide characteristic features of reflectance profiles in the lateral dimension, which may vary over different retinal regions and subject to affect the AV classification performance. Therefore, it is desirable to explore additional characteristic AV features in axial direction for performance improvement of AV classification. Recent efforts have been reported with the aim of improving AV characterization using the vessel wall information [33-35]. Kim et al. demonstrated AV classification using cross-sectional vascular features in OCT/OCTA of the mouse retina [33, 36]. Afsharan et al. observed a higher polarization score in artery vessel walls compared to that in veins which may imply different muscle tissue concentration in arteries and veins [34]. This study is to characterize OCT reflectance profiles of retinal blood vessels, and to test the potential of using these vascular features to guide artery-vein classification in OCTA of the human retina.

## 2. Materials and methods

### 2.1 Data acquisition

This study was approved by the Institutional Review Board of the University of Illinois at Chicago and was in pursuance with the ethical standards stated in the Declaration of Helsinki. Six healthy young subjects without a history of ocular and systemic disease ranging in age from 24 to 39 years were recruited from the Lions of Illinois Eye Research Institute of the University of Illinois. The mean refractive error of subjects was measured using ARK-900 autorefractor system (Nidek, San Jose, CA, USA) and found to be -2 ± 0.86 diopters (range: -3.25 to -1.75 diopters). Informed consent was provided by each subject before participation in the study.

AngioVue SD-OCT angiography system (Optovue, Fremont, CA, USA) with a 70-KHz A-scan rate was used to acquire OCT and OCTA images. The axial and lateral resolution was 5 μm and 15 μm, respectively. Each OCT/OCTA image corresponds to a 6 mm × 6 mm field of view (FOV). For OCTA reconstruction, split-spectrum amplitude-decorrelation angiography (SSADA) algorithm was employed [37], and the repeated 2 B-scans were averaged for OCT volume construction. OCT/OCTA volumetric images were exported from Optovue graphical user interface (GUI) and further reconstructed using custom-developed MATLAB R2021a (Mathworks, Natick, MA, USA) functions.

Color fundus images were captured using Pictor Plus (Volk Optical, Mentor, OH, USA), a nonmydriatic retinal camera with a FOV up to 45 degrees, with frame resolution of 1536 x 1152 pixels. The color-fundus images were exported from Pictor Plus and used to verify the AV classification in OCT/OCTA.

### 2.2 Data preparation and analysis

To obtain the cross-sectional vessel images, the OCT volumes (see Fig. 1A) were resliced by tracing individual blood vessels as shown in Fig. 1B. To obtain a single representative vessel image with enhanced contrast, several resliced B-scans at the center of the blood vessel were averaged (Fig. 1C, 1D). This image segmentation was performed in Fiji software [38], and applied to major vessel branches, directly emanating from the optic nerve head (ONH). For quantitative analysis, the cross-sectional vessel image (Fig. 1D) was further flattened, followed by a vertical rescaling that made the vessel images either down-sampled or up-sampled to ensure equal footing between different vessel images. The A-lines in each vessel image were then averaged in the lateral direction to produce the axial profile intensity of each vessel. After that, the intensity of all pixels was divided by the maximum pixel intensity for intensity normalization. Detailed procedure has been reported in a recent OCT study of mouse retinas [33].

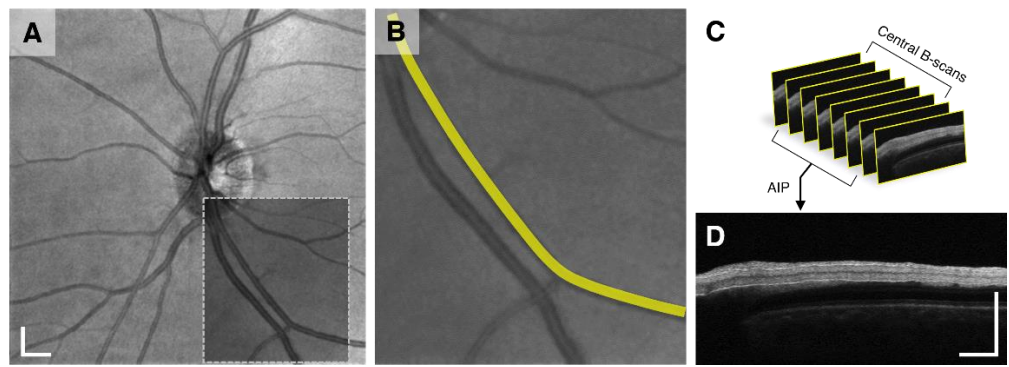

Fig. 1. Cross-sectional vessel image preparation. (A) Representative en face OCT. (B) Illustration of the vessel tracing process from the ONH to obtain a cross-sectional vessel image. The image is an enlarged view of the gray box in (A). The yellow line in (B) illustrates the maximum width from the center of the vessel used to obtain the B-scans in (C). (C) Stack of B-scans resliced from the vessel tracking in (B). AIP: average intensity projection. (D) Representative cross-sectional vessel image. Scale bars: 500 μm.

## 3. Results

### 3.1 Establishing ground truth for AV analysis

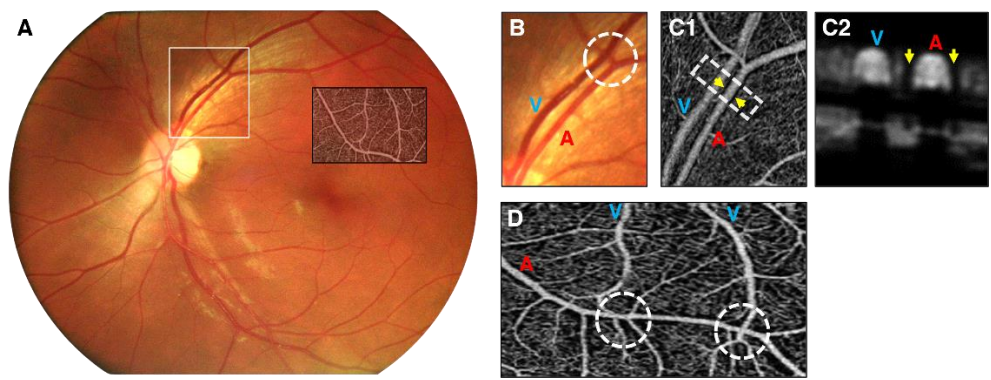

Fig. 2. Establishing ground truths for AV classification. (A) Representative fundus image. (B) Enlarged illustration of the white window region in (A). The white circle showing nearly 90-degree angle crossing between AV. In addition, the artery is brighter than the vein. (C1) Corresponding OCTA from the same region in (B). Yellow arrows indicate the capillary-free zone around the artery. (C2) Cross-sectional view of white box in (C1). Yellow arrows indicate the capillary-free zone around the artery. (D) Representative *en face* OCTA image. White circles show AV cross each other. The image is an enlarged view of the OCTA inset in (A). A: artery; V: vein.

The ground truth information was first constructed to separate cross-sectional vessel images into artery and vein for qualitative profile feature analysis. Color fundus photography and *en face* OCT/OCTA were utilized to determine ground truth of arteries and veins by manually examining established criteria for AV classification in retinal images (Fig. 2) [31, 39]. In Fig. 2A, the light reflex of the inner parts of the vessels was distinct in arteries. Arteries also exhibited a brighter color compared to veins due to the different oxygen level in the arteries and veins (Fig. 2B). En face OCTA further revealed a capillary-free zone around arteries (Fig. 2C1 and Fig. 2C2). In addition, artery and vein crossing formed a nearly 90-degree angle. We also confirmed that arteries did not cross arteries, and veins did not cross veins; while arteries cross veins and vice versa (Fig. 2D).

*3.2 Demographic analysis of retinal arteries and veins.*

OCT/OCTA dataset consisted of volumetric retinal images from six human subjects. Figure 3A and Fig. 3B show representative cross-sectional vessel images of arteries and veins. These cross-sectional vessel images were obtained at the first, second and third order branches respectively. The macular region contained blood vessels from the second and third branch only while the ONH contained mostly the first and few second branches. For analysis, 59 cross-sectional vessel images were obtained from the ONH, and 16 cross-sectional vessel images were obtained from the macula. In total, 75 cross-sectional vessel images were obtained from both the ONH and macula. The 75 vessel images were grouped into artery and vein for quantitative analysis based on the ground truth information. When comparing the venous and arterial sizes, we observed that the venous size varied more than the arterial size. The vein was the thickest in each of the subjects and showed more variation in sizes when compared with the arteries. We also confirmed the mean vessel width for the artery (91.7 µm ± 22.0 µm) and vein (124.89 µm ± 32.91 µm) (Fig. 3C) was consistent with a previous study [40].

*3.3 Comparative analysis of reflectance profiles in retinal arteries and veins*

To examine the reflectance profile along the vessel wall, we observed the cross-sectional vessel images in the different branches (Fig. 3A and Fig. 3B). Arteries reveal continuous hyper-reflective boundaries along the vessel walls. In subsequent branches, the hyper-reflective boundary intensity steadily decreases but remains distinct. However, hyper-reflectivity at the lower vessel boundary in veins was indiscernible in all branches at the current axial resolution, which might indicate thin or lack of vessel walls. To quantitatively analyze the different reflectance profiles, normalized vascular intensity maps were constructed from the different branch vessels from all subjects. Figure 3D and Fig. 3E show the average normalized intensity profiles from the intensity maps of arteries and veins. A sharp increase in the lower vessel boundary intensity was consistently observed in the arterial profile while a continuous decrease in the intensity with some pixels before the lower boundary was observed in the veins.

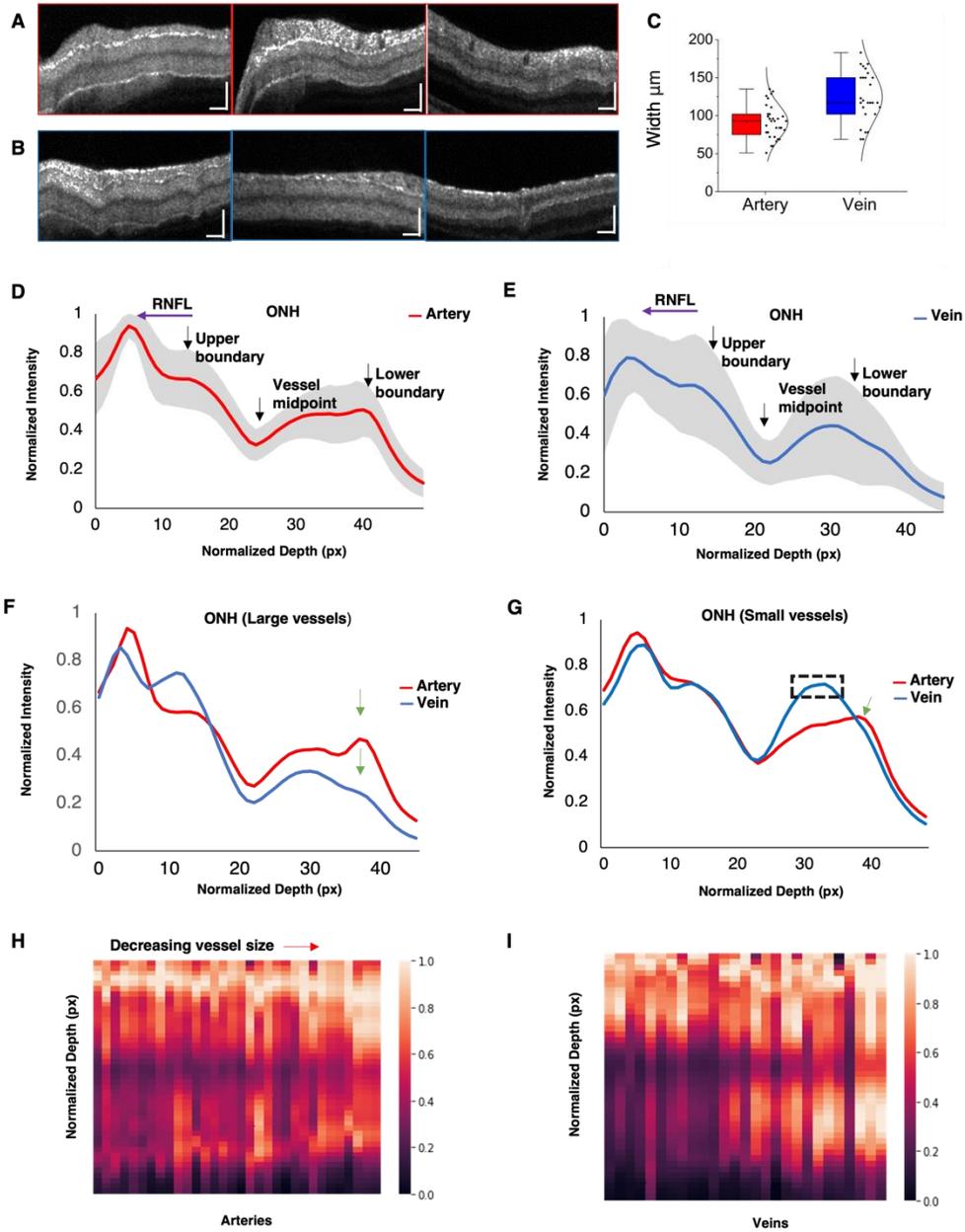

Fig. 3. Quantitative cross-sectional profile analysis for AV classification. Representative cross-sectional B-scans for (A) arteries and (B) veins at the first, second and third vessel branches. (C) Box-scatter plot of vessel width measurement from 6 subjects (N = 32 for arteries and N = 27 for veins). Average intensity plot constructed from (D) 32 arteries and (E) 27 veins. Gray areas show standard deviation. Overlaid average intensity plot constructed from (F) 16 large arteries and 14 large veins, and (G) 16 small arteries and 13 small veins. Green arrows indicate lower boundaries in (F) and lower boundary in (G). Black dotted box in (G) represents the hyper-reflective zone. Normalized intensity maps constructed from (H) 32 arteries and (I) 27 veins. Scale bars: 500 μm. RNFL: retinal nerve fiber layer.

### 3.4 Blood flow patterns in retinal arteries and veins

An interesting observation in this study was the layered intensity distribution in the venous lumen. To better understand these intensity patterns, blood vessels above the median size were categorized as large vessels while those below the median size were categorized as small vessels. The large vessels were obtained mostly at the first and second order branch while the small vessels were obtained at the second and third order branch. Figure 3F and Fig. 3G show an overlaid intensity profile of artery and vein with classification predicated on vessel size. From the hypo-reflective center of the lumen, the intensity distribution was observed to be relatively uniform in both large and small arteries. However, the intensity distribution of the vein was dependent on the vessel size as it tended to reveal a hyper-reflective zone at the bottom half of the lumen (below the hypo-reflective center) in small vessels. The AV intensity plots (see Fig. 3H and Fig. 3I) further revealed mostly homogenous intensity distribution in arteries, hypo-reflectivity in the bottom half of the venous lumen in large veins and hyper-reflectivity in the bottom half of the venous lumen in small veins. These results confirmed the unique speckle patterns inside the lumens of arteries and veins.

### 3.5 AV classification in OCT images

The axial profile features observed in retinal arteries and veins can be used for AV classification in OCT images. The source nodes for each of the vessels at the ONH region was identified and the corresponding reflectance profile was obtained (Fig. 4(A1), Fig. 4(B1), Fig. 4(C1), Fig. 4(D1) and Fig. 4(E1)). Consequently, blood vessel tracking algorithm from the source nodes [41], which employs curvature angle information was used to classify the entirety of the vasculature into artery or vein. However, due to limited spatial resolution, small arterioles and venules from the ONH did not clearly reveal the axial profile features for AV classification (Fig. 4(A2), Fig. 4(B2), Fig. 4(C2), Fig. 4(D2) and Fig. 4(E2)).

### 3.6 OCT feature analysis guided AV classification in OCTA

To guide AV classification in OCTA, the *en face* OCT vessel map was superimposed with the OCTA vessel map. To generate the OCTA vessel map, the outer retina was removed to clean small capillary mesh that cannot be resolved as artery or vein. The *en face* OCT and OCTA images were created by the three-dimensional projection of OCT B-scans and OCTA B-scans. Because the OCTA is naturally based on the OCT speckle variance processing to enhance the visibility of blood vessels, image registration is not required during superimposition. As a result, the *en face* OCT vessel map coordinate information is transferred to OCTA vessel map. Owing to additional vascular details contained in the OCTA vessel map, the endpoints and matching branch points are detected using morphological functions in MATLAB. The endpoints are then backtracked to the artery-vein branches to guide the AV classification and vessel tracking in OCTA (Fig. 4A3 – 4E3).

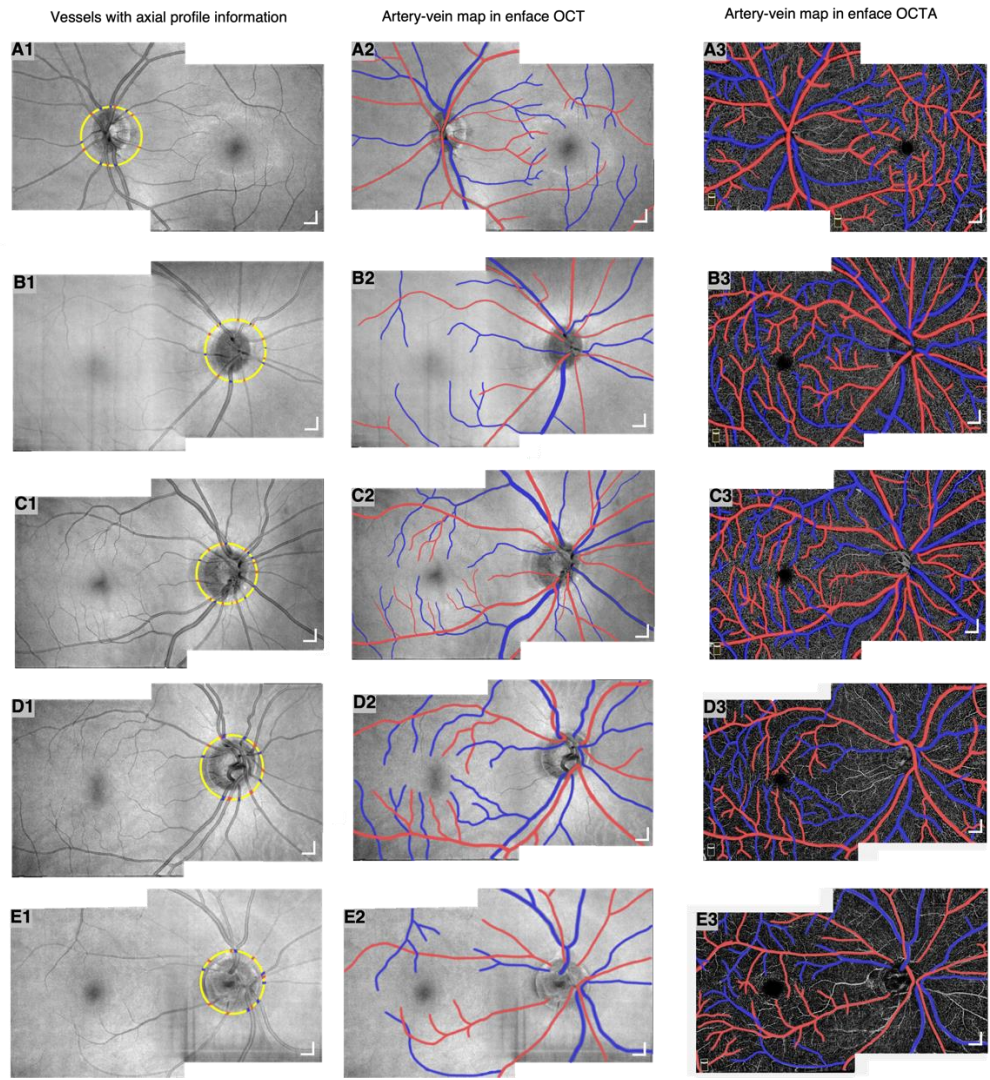

Fig. 4. Representative pseudo-colored AV vessel map in OCT and OCTA from five different subjects. Images centered at the ONH and macula were merged to form a montage over a 10 mm x 6 mm retinal area showing arteries (red) and veins (blue). The yellow circle in (A1, B1, C1, D1, E1) show source nodes of vessels with axial profile information. Corresponding OCT (A2, B2, C2, D2, E2) and OCTA (A3, B3, C3, D3, E3) montage images with AV vessels segmented. Scale bars: 500 μm.

## 4. Discussion

In summary, we characterize OCT axial reflectance profiles in retinal vessels, and validated the feasibility of using cross-sectional vascular analysis to guide AV classification in OCTA. Cross-sectional OCT vessel images were obtained from the ONH and macular regions. For characterizing OCT axial reflectance profiles, individual vessel images were first classified into artery and vein based on the established imaging features that appeared on fundus photographs and *en face* OCT/OCTA images. Normalized reflectance profiles revealed hyper-reflective wall boundaries in arteries while a layered intensity distribution in veins (Fig. 3).

The hyper-reflective boundary signals were evident in the arterial vessel walls. Prior histologic studies have described the different components of blood vessel walls. Blood vessel walls are made up of two types of cells: endothelial cells and mural cells. The inner wall lining is tiled with endothelial cells. The outside wall lining is tiled with mural cells. This is referred to as pericytes in smaller vessels while in bigger vessels it is referred to as vascular smooth muscle cells (SMCs) [42]. The retinal arteries are distinguishable from other organ arteries by their well-developed tunica media which is lined by layers of mural cells. The number of SMCs steadily decreases as the arteries branch into arterioles [43]. The venous wall, on the other hand, is relatively thin with only a single layer of endothelial cells and few SMCs [44]. Previous studies of rat retinal vessels also exhibited hyper-reflective boundary signals in arterial radial B-scans [33]. Thus, the arterial wall morphology would be responsible for the hyper-reflective boundary signals in arteries.

In addition, three unique luminal intensity distributions were observed in different vessel sizes. In small veins, a distinct hyper-reflective zone was observed in the bottom half of the lumen. Venous laminar flow may account for this hyper-reflective zone [45]. Due to laminar flow properties, axial mixing of blood cells leaving from the DCP would be restricted; thus, most erythrocytes may follow the basal blood stream of the lumen. However, when multiple branches of small venules merge to form a large venous branch, the laminar flow stream can be potentially disrupted at the merging point. In these larger veins, a hypo-reflective zone was observed in the bottom half of the lumen. We speculate that the large blood volume may be a primary cause of OCT signal attenuation. As a result, the bottom half of the lumen may appear dark due to light attenuation (Fig. 3I). In small and large arteries, we consistently observed relatively homogeneous intensity distribution with a hypo-reflective zone at the center of the lumen (see Fig. 3A). The observed hypo-reflective zone was mainly caused by shear-induced erythrocytes orientation in concert with the vessel boundaries which causes low backscattering of light [46].

Potential limitations of this study in clinical applications include limited spatial resolution of OCT volumetric scans in order to obtain the distinct reflectance profiles for small arterioles and venules around the ONH. For following studies, we anticipate incorporating the reflectance profile information with other imaging features [28] to improve the accuracy of AV classification in OCTA. Moreover, the OCT signal sensitivity depends on the incident angle of the illuminating light, relative to the blood vessel. However, except for the center of ONH, the orientation of most of blood vessels is nearly perpendicular to the incident light, and thus the OCT imaging performance of individual vessels are basically comparable. Furthermore, comparative study of reflectance profile features in normal and diseased eyes can be valuable to identify useful biomarkers for clinical management of retinopathies.

## 5. Conclusion

Depth-resolved OCT profile analysis disclosed characteristic features to differentiate arteries and veins. Retinal arteries were observed to have hyper-reflective boundaries at both upper and lower walls. In contrast, retinal veins showed hyper-reflectivity at the upper boundary only. Relatively uniform lumen intensity was observed in both small and large arteries. However, the vein lumen intensity was dependent on the vessel size. Small veins exhibited a hyper-reflective zone at the bottom half of the lumen, while large veins showed a hypo-reflective zone at the bottom half of the lumen. These cross-sectional characteristics enable reliable AV classification in OCT. In conjunction with blood vessel tracking, the OCT axial features can be used to guide AV analysis in OCTA to foster early diagnosis of retinal diseases. Moreover, further involvement of the axial features may improve the performance of deep learning based AV-Net for automated AV classification in OCTA [32].

**Funding**

National Eye Institute (R01 EY023522, R01 EY029673, R01 EY030101, R01 EY030842, P30 EY001792); Research to Prevent Blindness; Richard and Loan Hill Endowment.## Disclosures

The authors declare that there are no conflicts of interest related to this article.

## Data Availability

Data underlying the results presented in this paper are not publicly available at this time but may be obtained from the authors upon reasonable request.